\documentclass[10pt,conference]{IEEEtran}

\usepackage{graphicx}
\usepackage{subfig}
\usepackage{booktabs}
\usepackage{subfig}
\usepackage{listings}
\usepackage{multirow}
\usepackage{tabularx}
\usepackage{color}
\usepackage{multirow}
\usepackage{url}
\usepackage{array}
\usepackage{ctable}
\usepackage{amsmath}
\usepackage{algpseudocode}

\newcommand{\code}[1]{{\small\textsf{#1}}}

\newcommand{\model}{SAM}

\author{\IEEEauthorblockN{Tam The Nguyen,
Phong Minh Vu,
Tung Thanh Nguyen}

\IEEEauthorblockA{Department of Computer Science and Software Engineering\\
Auburn University\\
\{tam,lenniel,tung\}@auburn.edu}\\
}

\begin{document}
\title{API Misuse Correction: A Statistical Approach}
\maketitle

\begin{abstract}
Modern software development relies heavily on Application Programming Interface (API) libraries. However, there are often certain constraints on using API elements in such libraries. Failing to follow such constraints (API misuse) could lead to serious programming errors. Many approaches have been proposed to detect API misuses, but they still have low accuracy and cannot repair the detected misuses. In this paper, we propose SAM, a novel approach to detect and repair API misuses automatically. SAM uses statistical models to describe five factors involving in any API method call: related method calls, exceptions, pre-conditions, post-conditions, and values of arguments. These statistical models are trained from a large repository of high-quality production code. Then, given a piece of code, SAM verifies each of its method calls with the trained statistical models. If a factor has a sufficiently low probability, the corresponding call is considered as an API misuse. SAM performs an optimal search for editing operations to apply on the code until it has no API issue. 
\end{abstract} %
\section{Introduction}

%ok
In modern software development, programmers often rely heavily on Application Programming Interface (API) frameworks and libraries to shorten time-to-market and upgrade cycles of software systems. For example, a prior study reports some Android mobile apps having up to 42\% of their external dependencies to Android API and 68\% to Java API~\cite{syer_1}. However, programmers lacking programming experiences and relevant documentation (including code examples) could use API incorrectly (API misuse). For example, one can forget to check that \code{hasNext()} returns true before calling \code{next()} with an \code{Iterator} from Java API.

% \begin{figure}[h]
% 	\centering
% 	\includegraphics[scale = 0.45]{figures/Example2}
%     \caption{An example of an API misuse and the fix}
% 	\label{example5}
% \end{figure}

API misuse is a popular root cause of software errors, crashes, and vulnerabilities~\cite{Fahl-1,Monperrus_1,Sushin_1,Egele-1,Nadi,Georgiev,Amann_mubench, Tam}. Thus, researchers have proposed and developed several API misuse detectors. Unfortunately, recent studies show that those detectors often have low accuracy~\cite{Amann_system}. In addition, to the best of our knowledge, no current approaches can repair the detected API misuses automatically.

To understand the nature of API misuses, we studied a dataset of 144 API misuses provided by Amann \textit{et al.}~\cite{Amann_mubench} publicly. We found that the root causes of those API misuses can be classified into five groups including incorrect temporal order of method calls, incorrect handling of exceptions, missing pre-conditions and post-conditions, and incorrect values of arguments. Table~\ref{result} shows the total number of each kind of misuses in the dataset.

\begin{table}[h]
	\centering
	\sf
	\caption{Root causes of 144 API misuses}
	\label{result}
	\begin{tabular}{lr}
		\toprule
		Incorrect temporal order of method calls                    & 75     \\
		Incorrect handling of exceptions  & 27                          \\
		Missing pre-condition              & 22             \\
		Missing post-condition   & 15                      \\
		Incorrect argument value         & 5                \\
		\bottomrule
	\end{tabular}
	\vspace{-7pt}
\end{table}

\section{Approach}

In this paper, we propose {\model} \textit{(``\underline{S}tatistical Approach for \underline{A}PI \underline{M}isuses'')}, a statistical approach for detecting and repairing API misuses automatically. Suggested by Table 1, {\model} uses several statistical models for five factors involving the usage of any API method call: the temporal order between method calls, exception handling, pre-condition, post-condition, and argument values. {\model} trains those statistical models using a massive amount of high-quality production code available in open-source repositories and app stores.

After training, {\model}, can detect and repair API misuses in a given code snippet. It first validate every method call in such code using its trained statistical models. If one usage factor of a method call $m$ has a sufficiently low probability (e.g. less than a threshold), it considers the call as an API misuses. {\model} then consider the API misuse correction problem into an optimal search problem in which it searches for repair actions that optimally eliminate those low probability usage factors. 

\begin{figure}[h]
	\centering
	\includegraphics[scale = 0.5]{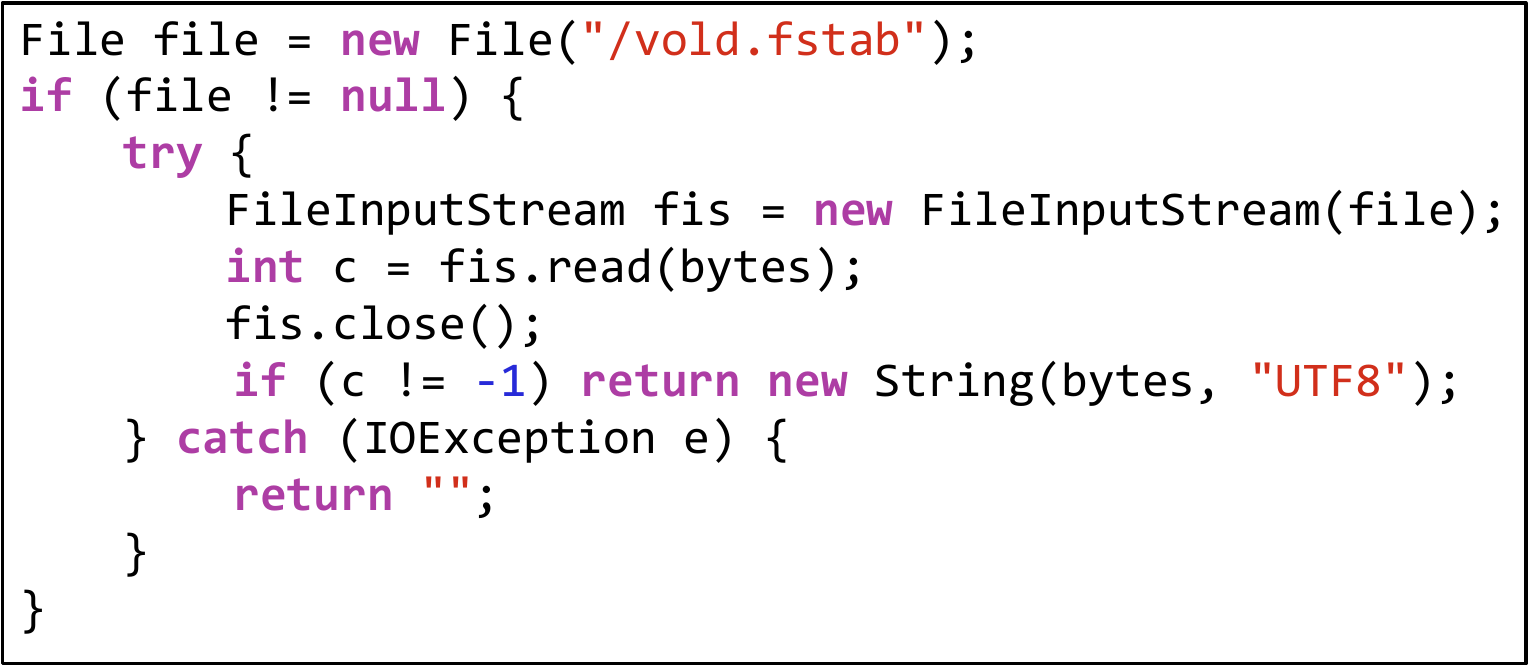}
    \caption{Reading a file with \code{File} and \code{FileOutputStream}}
	\label{example1}
	\vspace{-7pt}
\end{figure}

To demonstrate the model, Figure \ref{example1} shows a code example of API usage that involves 2 Java API objects \code{java.io.File} and \code{java.io.FileOutputStream}. We define five types of usage factors that represent the usages of API method calls and label them as follows.\\
$\bullet$ \textit{Temporal Order}: The temporal order specifies the order constraints between API method calls. In particular, the temporal order factor of an API method call \code{m} is defined as the API method call appears before \code{m} in the API usage. For example, the method call \code{fis.read(bytes)} in Figure \ref{example1} requires the initialization of the \code{FileInputStream} variable \code{fis}. $P_{\alpha}(m|C)$ is the probability of the temporal order factor of $m$ given the code context $C$, or $P(m_i|m)$. The method $m_i$ appears right before $m_i$ in the code context. \\
$\bullet$ \textit{Precondition}: The precondition factors of an API method call \code{m} are defined as all the preconditions that need to be checked on the calling object or parameters before calling \code{m}. For instance, before initialize the \code{FileInputStream} variable \code{fis}, the parameter \code{file} needs a \code{null} check. $P_{\beta}(\{m,p\}|C)$ is the precondition probability of the parameter $p$ in the method call $m$ in $C$. Each parameter has its own precondition probability, thus, $P_{\beta}(m|C)$ is the product of the those probabilities.\\ 
$\bullet$ \textit{Postcondition}: The postcondition factors of an API method call \code{m} are defined as all the postconditions that need to be checked on the calling object, parameters, or the returning value after calling \code{m}. For instance, after reading a character from the file using the \code{fis.read(bytes)} method, the return value is compared with a constant. $P_{\gamma}(m|C)$ represents of probability of postcondition factors of the method call $m$. \\
$\bullet$ \textit{Argument Value}: The argument value factor specifies the value of an argument when it is passed to an API method call. The argument value is limited by the type of argument and the specification of the method call. For example, in Figure \ref{example1}, the charset used in the string constructor is \code{UTF8}. If the argument value is not the name of a supported charset, the method call will throw a \code{UnsupportedEncodingException}. $P_{\delta}(m|C)$ represents of probability of argument value factors of the method call $m$. \\
$\bullet$ \textit{Exception}: The exception factors of an API method call \code{m} are defined as the exceptions that need to be handled when calling \code{m}. For instance, when calling \code{fis.read(bytes)}, it is required to handle \code{IOException} exception. $P_{\epsilon}(m|C)$ represents of probability of exception factors of the method call $m$.

{\model} represents the usage of an API method call $m$ given the context $C$ as the combination of its usage factors. Given {\model}, we could detect API misuses in code. An API method call is considered a misuse if there is at least one usage factor of a method $m$ has a sufficiently low probability (e.g. less than a threshold). Thus, the API misuse detector simply checks all probabilities of usage factors to detect API misuses. 

The API misuse correction problem is modeled as an optimal search problem. The algorithm to solve the problem is described in Figure \ref{algo1}. The input of the algorithm includes the code that contains the API misuse(s), and the current editing length $L$. In line 2, the function \Call{Detect-API-Misuses}{$C$} finds all the usage factors $X$ that leads to the API misuse(s). If $X = \emptyset$, the correction is finished and $C$ is returned. Next, the algorithm compares the current editing length $L$ with the maximum editing length $MaxLength$ in line 6. It then generates repair actions $A$ based on $X$ with the function \Call{Generate-Repair-Actions}{$C$}. Each repair action $a$ is applied on $C$ in line 9. The function \Call{Correct-API-Misuses}{$C', L+1$} is called recursively to further edit the code by one edit length. The algorithm stops when there is no API misuse(s) in the code or the maximum editing length $MaxLength$ is reached.

\begin{figure}  
\scriptsize
\begin{algorithmic}[1]  
\Function{Correct-API-Misuses}{Code $C$, EditLength $L$}
\State $X \gets$ \Call{Detect-API-Misuses}{C}
\If{$X = \emptyset$}
\State \textbf{return} $C$
\EndIf
\If{$L < MaxLength$ }
\State $A\gets$ \Call{Generate-Repair-Actions}{X} 
\For{Action $a$ in $A$}
\State $C'\gets$ \Call{Modify}{$C, a$}  
\State \Call{Correct-API-Misuses}{$C', L+1$}
\EndFor
\EndIf
\EndFunction  
\end{algorithmic}  
\caption{The API misuse correction algorithm}  
\label{algo1}  
\vspace{-7pt}
\end{figure}

Training {\model} requires calculating the probability distribution of each usage factor of every API method call $m$. For example, the probability distribution of the temporal order factor of the method call $m$ requires computing all probabilities $P(m_i|m)$, where the method $m$ appears right before $m$. We have,
\begin{equation}
\scriptsize
    P(m_i|m) = \frac{P(m_{i}m)}{P(m)} \approx \frac{N(m_{i}m)+1}{N(m)+1} \nonumber
\end{equation}
where $N(m_{i}m)$ is number of occurrences of the bigram $m_{i}m$. $N(m)$ is the number of occurrences of $m$. Thus, the training of {\model} involves counting those occurrences. This is an advantage of {\model} as it could easily be trained from massive amount of code to improve the accuracy.

\section{Related Work}
The researchers have proposed several API-misuse detectors including~\cite{Jadet, GROUMINER, DMMC, TIKANGA,PR_Miner,RGJ07,CHRONICLER,AX09}. Jadet~\cite{Jadet} and GroumMiner~\cite{GROUMINER} detect API misuses by detecting violation on object usage graphs. DMMC~\cite{DMMC} is a misuse detector for Java specialized in missing method calls. Tikanga~\cite{TIKANGA} is a misuse detector for Java that builds on JADET~\cite{Jadet}. Amann \textit{et al.}~\cite{Amann_system} provides a systematic evaluation of static API misuse detectors. They also provided MuBench~\cite{Amann_mubench}, a benchmark for evaluating API misuse detectors. They proposed MuDetect~\cite{Amann_mudetech}, an improved API misuse detector of GroumMiner~\cite{GROUMINER}.

% There are detectors for designed for programming language C. PrMiner~\cite{PR_Miner} uses frequent-itemset mining to detect missing method calls. RGJ07~\cite{RGJ07} detects missing call-argument conditions by using frequent itemset mining. Chronicler~\cite{CHRONICLER} detects call-order violations by mining frequent call-precedence relations. AX09~\cite{AX09} detects missing error handling by push-down model checking.

Car-Miner~\cite{CAR_MINER} is a wrong error handling detector for C++ and Java. ExAssist~\cite{ExAssist} is a tool for detecting exception handling bugs and automatically fix those bugs. Alattin~\cite{ALATTIN} is a detector for Java that detects missing null checks, missing value or state conditions not involving literals, and missing calls required in
checks. DroidAssist~\cite{DroidAssist} is a detector for Dalvik Bytecode. MAD-API~\cite{MAD-API} is an approach that fixes Android APIs that are out of date.

%It uses a Hidden Markov Model to compute the likelihood of call sequences to detect missing, misplaced, and redundant method calls.
\section{Conclusions}
In this paper, we propose {\model}, a statistical approach for detecting and repairing API misuses automatically. Given a piece of code, SAM verifies each of its method calls with the trained statistical models. If a factor has a sufficiently low probability, the corresponding call is considered as an API misuse. SAM performs an optimal search for editing operations to apply on the code until it has no API issue.

\bibliographystyle{IEEEtran}
\bibliography{IEEEabrv,apicorrection}

% Generated by IEEEtran.bst, version: 1.12 (2007/01/11)
\begin{thebibliography}{10}
\providecommand{\url}[1]{#1}
\csname url@samestyle\endcsname
\providecommand{\newblock}{\relax}
\providecommand{\bibinfo}[2]{#2}
\providecommand{\BIBentrySTDinterwordspacing}{\spaceskip=0pt\relax}
\providecommand{\BIBentryALTinterwordstretchfactor}{4}
\providecommand{\BIBentryALTinterwordspacing}{\spaceskip=\fontdimen2\font plus
\BIBentryALTinterwordstretchfactor\fontdimen3\font minus
  \fontdimen4\font\relax}
\providecommand{\BIBforeignlanguage}[2]{{%
\expandafter\ifx\csname l@#1\endcsname\relax
\typeout{** WARNING: IEEEtran.bst: No hyphenation pattern has been}%
\typeout{** loaded for the language `#1'. Using the pattern for}%
\typeout{** the default language instead.}%
\else
\language=\csname l@#1\endcsname
\fi
#2}}
\providecommand{\BIBdecl}{\relax}
\BIBdecl

\bibitem{syer_1}
\BIBentryALTinterwordspacing
M.~D. Syer, M.~Nagappan, A.~E. Hassan, and B.~Adams, ``Revisiting prior
  empirical findings for mobile apps: An empirical case study on the 15 most
  popular open-source android apps,'' in \emph{Proceedings of the 2013
  Conference of the Center for Advanced Studies on Collaborative Research},
  ser. CASCON '13.\hskip 1em plus 0.5em minus 0.4em\relax Riverton, NJ, USA:
  IBM Corp., 2013, pp. 283--297. [Online]. Available:
  \url{http://dl.acm.org/citation.cfm?id=2555523.2555553}
\BIBentrySTDinterwordspacing

\bibitem{Fahl-1}
\BIBentryALTinterwordspacing
S.~Fahl, M.~Harbach, T.~Muders, L.~Baumg\"{a}rtner, B.~Freisleben, and
  M.~Smith, ``Why eve and mallory love android: An analysis of android ssl
  (in)security,'' in \emph{Proceedings of the 2012 ACM Conference on Computer
  and Communications Security}, ser. CCS '12.\hskip 1em plus 0.5em minus
  0.4em\relax New York, NY, USA: ACM, 2012, pp. 50--61. [Online]. Available:
  \url{http://doi.acm.org/10.1145/2382196.2382205}
\BIBentrySTDinterwordspacing

\bibitem{Monperrus_1}
\BIBentryALTinterwordspacing
M.~Monperrus and M.~Mezini, ``Detecting missing method calls as violations of
  the majority rule,'' \emph{ACM Trans. Softw. Eng. Methodol.}, vol.~22, no.~1,
  pp. 7:1--7:25, Mar. 2013. [Online]. Available:
  \url{http://doi.acm.org/10.1145/2430536.2430541}
\BIBentrySTDinterwordspacing

\bibitem{Sushin_1}
\BIBentryALTinterwordspacing
J.~Sushine, J.~D. Herbsleb, and J.~Aldrich, ``Searching the state space: A
  qualitative study of api protocol usability,'' in \emph{Proceedings of the
  2015 IEEE 23rd International Conference on Program Comprehension}, ser. ICPC
  '15.\hskip 1em plus 0.5em minus 0.4em\relax Piscataway, NJ, USA: IEEE Press,
  2015, pp. 82--93. [Online]. Available:
  \url{http://dl.acm.org/citation.cfm?id=2820282.2820295}
\BIBentrySTDinterwordspacing

\bibitem{Egele-1}
\BIBentryALTinterwordspacing
M.~Egele, D.~Brumley, Y.~Fratantonio, and C.~Kruegel, ``An empirical study of
  cryptographic misuse in android applications,'' in \emph{Proceedings of the
  2013 ACM SIGSAC Conference on Computer \&\#38; Communications Security}, ser.
  CCS '13.\hskip 1em plus 0.5em minus 0.4em\relax New York, NY, USA: ACM, 2013,
  pp. 73--84. [Online]. Available:
  \url{http://doi.acm.org/10.1145/2508859.2516693}
\BIBentrySTDinterwordspacing

\bibitem{Nadi}
S.~{Nadi}, S.~{Krüger}, M.~{Mezini}, and E.~{Bodden}, ``"jumping through
  hoops": Why do java developers struggle with cryptography apis?'' in
  \emph{2016 IEEE/ACM 38th International Conference on Software Engineering
  (ICSE)}, May 2016, pp. 935--946.

\bibitem{Georgiev}
\BIBentryALTinterwordspacing
M.~Georgiev, S.~Iyengar, S.~Jana, R.~Anubhai, D.~Boneh, and V.~Shmatikov, ``The
  most dangerous code in the world: Validating ssl certificates in non-browser
  software,'' in \emph{Proceedings of the 2012 ACM Conference on Computer and
  Communications Security}, ser. CCS '12.\hskip 1em plus 0.5em minus
  0.4em\relax New York, NY, USA: ACM, 2012, pp. 38--49. [Online]. Available:
  \url{http://doi.acm.org/10.1145/2382196.2382204}
\BIBentrySTDinterwordspacing

\bibitem{Amann_mubench}
S.~{Amann}, S.~{Nadi}, H.~A. {Nguyen}, T.~N. {Nguyen}, and M.~{Mezini},
  ``Mubench: A benchmark for api-misuse detectors,'' in \emph{2016 IEEE/ACM
  13th Working Conference on Mining Software Repositories (MSR)}, May 2016, pp.
  464--467.

\bibitem{Tam}
\BIBentryALTinterwordspacing
T.~T. Nguyen, P.~M. Vu, and T.~T. Nguyen, ``An empirical study of exception
  handling bugs and fixes,'' in \emph{Proceedings of the 2019 ACM Southeast
  Conference}, ser. ACM SE '19.\hskip 1em plus 0.5em minus 0.4em\relax New
  York, NY, USA: ACM, 2019, pp. 257--260. [Online]. Available:
  \url{http://doi.acm.org/10.1145/3299815.3314472}
\BIBentrySTDinterwordspacing

\bibitem{Amann_system}
S.~{Amann}, H.~A. {Nguyen}, S.~{Nadi}, T.~N. {Nguyen}, and M.~{Mezini}, ``A
  systematic evaluation of static api-misuse detectors,'' \emph{IEEE
  Transactions on Software Engineering}, pp. 1--1, 2018.

\bibitem{Jadet}
\BIBentryALTinterwordspacing
A.~Wasylkowski, A.~Zeller, and C.~Lindig, ``Detecting object usage anomalies,''
  in \emph{Proceedings of the the 6th Joint Meeting of the European Software
  Engineering Conference and the ACM SIGSOFT Symposium on The Foundations of
  Software Engineering}, ser. ESEC-FSE '07.\hskip 1em plus 0.5em minus
  0.4em\relax New York, NY, USA: ACM, 2007, pp. 35--44. [Online]. Available:
  \url{http://doi.acm.org/10.1145/1287624.1287632}
\BIBentrySTDinterwordspacing

\bibitem{GROUMINER}
\BIBentryALTinterwordspacing
T.~T. Nguyen, H.~A. Nguyen, N.~H. Pham, J.~M. Al-Kofahi, and T.~N. Nguyen,
  ``Graph-based mining of multiple object usage patterns,'' in
  \emph{Proceedings of the the 7th Joint Meeting of the European Software
  Engineering Conference and the ACM SIGSOFT Symposium on The Foundations of
  Software Engineering}, ser. ESEC/FSE '09.\hskip 1em plus 0.5em minus
  0.4em\relax New York, NY, USA: ACM, 2009, pp. 383--392. [Online]. Available:
  \url{http://doi.acm.org/10.1145/1595696.1595767}
\BIBentrySTDinterwordspacing

\bibitem{DMMC}
M.~Monperrus, M.~Bruch, and M.~Mezini, ``Detecting missing method calls in
  object-oriented software,'' in \emph{ECOOP 2010 -- Object-Oriented
  Programming}, T.~D'Hondt, Ed.\hskip 1em plus 0.5em minus 0.4em\relax Berlin,
  Heidelberg: Springer Berlin Heidelberg, 2010, pp. 2--25.

\bibitem{TIKANGA}
\BIBentryALTinterwordspacing
A.~Wasylkowski and A.~Zeller, ``Mining temporal specifications from object
  usage,'' in \emph{Proceedings of the 2009 IEEE/ACM International Conference
  on Automated Software Engineering}, ser. ASE '09.\hskip 1em plus 0.5em minus
  0.4em\relax Washington, DC, USA: IEEE Computer Society, 2009, pp. 295--306.
  [Online]. Available: \url{https://doi.org/10.1109/ASE.2009.30}
\BIBentrySTDinterwordspacing

\bibitem{PR_Miner}
\BIBentryALTinterwordspacing
Z.~Li and Y.~Zhou, ``Pr-miner: Automatically extracting implicit programming
  rules and detecting violations in large software code,'' in \emph{Proceedings
  of the 10th European Software Engineering Conference Held Jointly with 13th
  ACM SIGSOFT International Symposium on Foundations of Software Engineering},
  ser. ESEC/FSE-13.\hskip 1em plus 0.5em minus 0.4em\relax New York, NY, USA:
  ACM, 2005, pp. 306--315. [Online]. Available:
  \url{http://doi.acm.org/10.1145/1081706.1081755}
\BIBentrySTDinterwordspacing

\bibitem{RGJ07}
\BIBentryALTinterwordspacing
M.~K. Ramanathan, A.~Grama, and S.~Jagannathan, ``Static specification
  inference using predicate mining,'' in \emph{Proceedings of the 28th ACM
  SIGPLAN Conference on Programming Language Design and Implementation}, ser.
  PLDI '07.\hskip 1em plus 0.5em minus 0.4em\relax New York, NY, USA: ACM,
  2007, pp. 123--134. [Online]. Available:
  \url{http://doi.acm.org/10.1145/1250734.1250749}
\BIBentrySTDinterwordspacing

\bibitem{CHRONICLER}
M.~K. {Ramanathan}, A.~{Grama}, and S.~{Jagannathan}, ``Path-sensitive
  inference of function precedence protocols,'' in \emph{29th International
  Conference on Software Engineering (ICSE'07)}, May 2007, pp. 240--250.

\bibitem{AX09}
\BIBentryALTinterwordspacing
M.~Acharya and T.~Xie, ``Mining api error-handling specifications from source
  code,'' in \emph{Proceedings of the 12th International Conference on
  Fundamental Approaches to Software Engineering: Held As Part of the Joint
  European Conferences on Theory and Practice of Software, ETAPS 2009}, ser.
  FASE '09.\hskip 1em plus 0.5em minus 0.4em\relax Berlin, Heidelberg:
  Springer-Verlag, 2009, pp. 370--384. [Online]. Available:
  \url{http://dx.doi.org/10.1007/978-3-642-00593-0_25}
\BIBentrySTDinterwordspacing

\bibitem{Amann_mudetech}
\BIBentryALTinterwordspacing
S.~Amann, H.~A. Nguyen, S.~Nadi, T.~N. Nguyen, and M.~Mezini, ``Investigating
  next steps in static api-misuse detection,'' in \emph{Proceedings of the 16th
  International Conference on Mining Software Repositories}, ser. MSR
  '19.\hskip 1em plus 0.5em minus 0.4em\relax Piscataway, NJ, USA: IEEE Press,
  2019, pp. 265--275. [Online]. Available:
  \url{https://doi.org/10.1109/MSR.2019.00053}
\BIBentrySTDinterwordspacing

\bibitem{CAR_MINER}
\BIBentryALTinterwordspacing
S.~Thummalapenta and T.~Xie, ``Mining exception-handling rules as sequence
  association rules,'' in \emph{Proceedings of the 31st International
  Conference on Software Engineering}, ser. ICSE '09.\hskip 1em plus 0.5em
  minus 0.4em\relax Washington, DC, USA: IEEE Computer Society, 2009, pp.
  496--506. [Online]. Available:
  \url{http://dx.doi.org/10.1109/ICSE.2009.5070548}
\BIBentrySTDinterwordspacing

\bibitem{ExAssist}
T.~T. Nguyen, P.~M. Vu, and T.~T. Nguyen, ``Recommending exception handling
  code,'' in \emph{ICSME}, 2019.

\bibitem{ALATTIN}
S.~{Thummalapenta} and T.~{Xie}, ``Alattin: Mining alternative patterns for
  detecting neglected conditions,'' in \emph{2009 IEEE/ACM International
  Conference on Automated Software Engineering}, Nov 2009, pp. 283--294.

\bibitem{DroidAssist}
T.~T. {Nguyen}, H.~V. {Pham}, P.~M. {Vu}, and T.~T. {Nguyen}, ``Recommending
  api usages for mobile apps with hidden markov model,'' in \emph{2015 30th
  IEEE/ACM International Conference on Automated Software Engineering (ASE)},
  Nov 2015, pp. 795--800.

\bibitem{MAD-API}
T.~Luo, J.~Wu, M.~Yang, S.~Zhao, Y.~Wu, and Y.~Wang, ``Mad-api: Detection,
  correction and explanation of api misuses in distributed android
  applications,'' in \emph{Artificial Intelligence and Mobile Services -- AIMS
  2018}, M.~Aiello, Y.~Yang, Y.~Zou, and L.-J. Zhang, Eds.\hskip 1em plus 0.5em
  minus 0.4em\relax Cham: Springer International Publishing, 2018, pp.
  123--140.

\end{thebibliography}

\end{document}